\documentclass{article}
\usepackage{epsfig}
\newcommand{\beq}{\begin{equation}}
\newcommand{\eeq}{\end{equation}}
\newcommand{\beqa}{\begin{eqnarray}}
\newcommand{\eeqa}{\end{eqnarray}}
\newcommand{\bea}{\begin{eqnarray}}
\newcommand{\eea}{\end{eqnarray}}

\newcommand   {\etal}    {{\it et~al.}}
\newcommand   {\bv}      {\hat b}
\newcommand   {\ev}[1]   {\langle #1\rangle}

\renewcommand   {\r}       {{\bf r}}

\newcommand  {\COSB}     {{\it Curr.\ Opin.\ Struct.\ Biol.\ }}
\newcommand  {\EL}       {{\it Europhys.\ Lett.\ }}
\newcommand  {\FD}       {{\it Fold.\ Des.\ }}

\newcommand  {\JCP}      {{\it J.\ Chem.\ Phys.\ }}
\newcommand  {\JMB}      {{\it J.\ Mol.\ Biol.\ }}
\newcommand  {\JP}       {{\it J.\ Phys.\ }}
\newcommand  {\JPC}      {{\it J.\ Phys.\ Chem.\ }}

\newcommand  {\M}        {{\it Macromolecules\ }}

\newcommand  {\ProEng}   {{\it Protein\ Eng.\ }}

\newcommand  {\PNAS}     {{\it Proc.\ Natl.\ Acad.\ Sci.\ USA\ }}
\newcommand  {\PR}       {{\it Phys.\ Rev.\ }}
\newcommand  {\PRL}      {{\it Phys.\ Rev.\ Lett.\ }}
\newcommand  {\Sci}      {{\it Science\ }}

\input{psfig}
\begin{document}

\begin{titlepage}

\begin{flushright}
LU TP 98-10\\
Revised Version\\
\today\\
\end{flushright}

\vspace{0.3in}

\LARGE
\begin{center}
{\bf Design of Sequences with Good Folding Properties\\ 
in Coarse-Grained Protein Models}\\
\vspace{.2in}
\large
Anders Irb\"ack\footnote{irback@thep.lu.se}, 
Carsten Peterson\footnote{carsten@thep.lu.se},\\ 
Frank Potthast\footnote{frank.potthast@hassle.se.astra.com}$^,$\footnote{Current address: 
Astra H\"assle AB, Preclinical R\&D, Research Informatics, 
S-431 83 M\"olndal, Sweden.} and 
Erik Sandelin\footnote{erik@thep.lu.se} \\
\vspace{0.10in}
Complex Systems Group, Department of Theoretical Physics\\ 
University of Lund,  S\"{o}lvegatan 14A,  S-223 62 Lund, Sweden \\
{\tt http://www.thep.lu.se/tf2/complex/} \\

\vspace{0.2in}  

Submitted to {\it Structure with Folding \& Design}

\end{center}

\normalsize

\vspace{0.1in}
\normalsize

{\bf Background:} Designing amino acid sequences that are stable
in a given target structure amounts to maximizing a conditional
probability. A straightforward approach to accomplish this
is a nested Monte Carlo where the conformation space is
explored over and over again for different fixed sequences,
which requires excessive computational demand. Several approximate
attempts to remedy this situation, based on energy
minimization for fixed structure or high-$T$ expansions,
have been proposed. These methods are fast but often not 
accurate since folding occurs at low $T$.

{\bf Results:} We develop a multisequence Monte Carlo procedure, 
where both sequence and conformation space are simultaneously 
probed with efficient prescriptions for pruning sequence space. 
The method is explored on hydrophobic/polar models.
We first discuss short lattice chains, in order to compare  
with exact data and with other methods. The method is then successfully 
applied to lattice chains with up to 50 monomers, and to off-lattice 20-mers. 

{\bf Conclusions:} The multisequence Monte Carlo method offers
a new approach to sequence design in coarse-grained models. 
It is much more efficient than previous Monte Carlo methods, 
and is, as it stands, applicable to a fairly wide range of 
two-letter models. 

Correspondence: Anders Irb\"ack\\
e-mail: irback@thep.lu.se


\end{titlepage}

\large

\normalsize

\section{Introduction}
\label{sec_intro}

The protein design problem amounts to finding an amino acid sequence 
given a target structure, which is stable in the target structure,
and is able to fold fast into this structure. In a typical model
the second requirement implies that stability must set in at not 
too low a temperature. Hence, one is led to consider the problem 
of finding sequences that maximize the stability of the target 
structure at a given temperature. In terms of a model described by 
an energy function $E(r,\sigma)$, where $r=\{\r_1,\r_2,..,\r_N\}$ 
denotes the structure coordinates and 
$\sigma=\{\sigma_1,\sigma_2,..,\sigma_N\}$ the amino acid sequence, 
this can be expressed as maximizing the conditional probability 
\beq
\label{P}
P(r_0|\sigma) = \frac{1}{Z(\sigma)}\exp [-E(r_0,\sigma)/T]\,,
\eeq
where $r_0$ denotes the target structure, $T$ the temperature 
and the partition function $Z(\sigma)$ is given by
\beq
\label{Z}
Z(\sigma) = \sum_r \exp[-E(r,\sigma)/T]\,.
\eeq
Maximizing $P(r_0|\sigma)$ with respect to $\sigma$ represents 
quite some challenge, since for any move in $\sigma$,  
the partition function $Z(\sigma)$ needs to be evaluated; each
evaluation of $P(r_0|\sigma)$ effectively amounts to a 
folding calculation for fixed sequence $\sigma$. 

Different ways of handling this sequence optimization problem 
have been proposed and partly explored in the context of 
coarse-grained (or minimalist) protein models, where amino acid 
residues represent the entities.
The proposed methods fall into three categories: 
\begin{itemize}
\item {\bf $E(r_0,\sigma)$-minimization} 
\cite{Shakhnovich:93a,Shakhnovich:93b,Shakhnovich:94}. 
If one simply ignores $Z(\sigma)$ in Eq.~(\ref{P}), one is left
with the problem of minimizing $E(r_0,\sigma)$. This is too crude, 
since for many coarse-grained models it implies that 
all $\sigma$ values line up to a homopolymer solution. This can be 
remedied by adding a constraint to $E(r_0,\sigma)$ restricting the 
overall composition. This method is very fast since no exploration of 
the conformation space is involved, but it does fail for a number 
of examples even for small system sizes.
\item {\bf High-$T$ expansion} 
\cite{Deutsch:95,Deutsch:96,Morrissey:96}. 
A more systematic approach is to approximate $Z(\sigma)$ with  
low-order terms in a cumulant or high-$T$ expansion. This method   
is also fast, and slightly more accurate than the $E(r_0,\sigma)$-minimization
method, but can also fail since folding takes place at low $T$.  
\item {\bf Nested MC (NMC)} \cite{Seno:96}. In order to avoid introducing 
uncontrolled approximations, one is forced to turn to Monte Carlo (MC) methods.
The most straightforward MC approach is to use a normal fixed-$\sigma$ MC in $r$ 
for estimating the $Z(\sigma)$ contribution to Eq.~(\ref{P}), which, however,
leads to a nested algorithm with a highly non-trivial inner part.  
Although correct results have been reported for toy-sized problems, 
this approach is inhibitorily CPU time-consuming for larger 
problem sizes.  
\end{itemize}
In this paper we develop and explore an alternative MC methodology, 
{\bf Multisequence (MS)} design, where the basic strategy is to 
create an enlarged configuration space; the sequence $\sigma$ becomes a 
dynamical variable~\cite{Irback:95b}. Hence, $r$ and $\sigma$ are 
put on a more equal footing, which, in particular, enables us to 
avoid a nested MC. Early stages of this project were reported 
in~\cite{Irback:97}.

The multisequence approach is explored on both a two-dimensional (2D) 
lattice model, the HP model of Lau and Dill~\cite{Lau:89}, and a 
simple three-dimensional (3D) off-lattice models~\cite{Irback:96c}, 
with very good results. As with any design method, one needs access
to suitable target structures, and also to verify the results by   
folding calculations. For short chains in the 2D HP model, 
$N \leq 18$, both these tasks are easy since all configurations 
can be enumerated. For longer lattice chains and off-lattice models 
powerful MC algorithms like simulated tempering 
\cite{Lyubartsev:92,Marinari:92,Irback:95b} are needed for the 
verification. 

Our calculations for the HP model can be divided into two
groups corresponding to short ($N=16$ and 18) and long ($N=32$ and 50)
chains. The results for short chains are compared to exact 
enumerations, and we find that 
our method reproduces the exact results extremely rapidly. 
We also compare our results to those obtained by   
$E(r_0,\sigma)$-minimization and a high-$T$ approach. 
It should be mentioned that for the former we 
scan through all possible fixed overall compositions, thereby giving this 
method a fair chance. Also, we make a detailed exact 
calculation illuminating the limitations of the high-$T$ expansion 
approach. 

For larger $N$ a ``bootstrap'' method is developed 
that overcomes the problem of keeping all possible 
sequences in the computer memory. The efficiency of this trick is 
illustrated for a $N=32$ target structure, which is chosen ``by hand''. 
Finally, a $N=50$ target structure is generated by using a
design algorithm that aims at throwing away those sequences 
that are unsuitable for {\it any} structure. This $N=50$ target 
structure is subsequently subject to our multisequence design approach, 
which readily finds a sequence with the target structure as its 
unique ground state. 
As a by-product, having access to good $N=50$ sequences, we 
investigate the behavior at the folding transition which, to our
knowledge, has not been studied before for comparable chain lengths.   

Earlier studies of the 3D off-lattice model~\cite{Irback:96c}, and a 
similar 2D model~\cite{Irback:96b}, have shown that the stability, 
as measured by the average size $\ev{\delta^2}$ of thermal structural 
fluctuations, is strongly sequence dependent. Here we perform
design experiments using native structures of both stable (low 
$\ev{\delta^2}$) and unstable (high $\ev{\delta^2}$) sequences 
as target structures. The quality of the designed sequences is 
carefully examined by monitoring the thermal average of the 
mean-square distance to the target structure, $\ev{\delta_0^2}$. 
We find that the method consistently improves on $\ev{\delta_0^2}$ 
and that it performs better than the $E(r_0,\sigma)$-minimization 
approach.   

This paper is organized as follows. Section~\ref{sec_method} contains our 
multisequence approach together with implementation issues. In 
Sec.~\ref{sec_lattice} the method is applied to the 2D HP model 
for sizes varying from $N=16$ to 50. In this section, also comparisons between 
the different approaches are performed. The efficiency of the multisequence 
method is discussed in Sec.~\ref{sec_mult}. In Sec.~\ref{sec_off-lattice} 
3D off-lattice model structures are designed, and 
Sec.~\ref{sec_summary} contains a brief summary.

\section{The Method}
\label{sec_method}

\subsection{Optimizing Conditional Probabilities}

Maximizing the conditional probability $P(r_0|\sigma)$ of Eq.~(\ref{P}) 
with respect to $\sigma$ for a given target structure $r_0$ is a challenge 
since it requires exploration of both conformation and sequence degrees 
of freedom. At high $T$ this task can be approached by using a 
cumulant expansion of $Z(\sigma)$, which makes the problem much easier.
Unfortunately, this is not the temperature regime of primary interest. 
In this paper we present an efficient MC-based procedure 
for sequence optimization at biologically relevant temperatures.

The problem of maximizing $P(r_0|\sigma)$ can be 
reformulated in terms of $P(\sigma|r_0)$ by introducing a marginal 
distribution of $\sigma$, $P(\sigma)$, and the corresponding joint 
distribution $P(r,\sigma)=P(r|\sigma)P(\sigma)$. Assigning equal a priori 
probability to all the $\sigma$, i.e. $P(\sigma)={\rm constant}$, one obtains
\beq
P(r_0|\sigma)=\frac{P(\sigma|r_0)P(r_0)}{P(\sigma)}\propto P(\sigma|r_0)\,,
\eeq
so maximizing $P(r_0|\sigma)$ is then equivalent to maximizing 
$P(\sigma|r_0)$.

In this paper we focus on the problem of designing a single structure
$r_0$. This is a special case of the more general problem of 
maximizing the probability 
\beq
 \sum_{r\in D} P(r|\sigma)
 \label{fuzzy}
\eeq
for a group of desired structures, $D$. Note that for a general set $D$ 
with more than one structure, this is not equivalent to maximizing 
$\sum_{r\in D}P(\sigma|r)$, since 
\beq
\label{noneq}
 \sum_{r\in D} P(r|\sigma)=\sum_{r\in D}\frac{P(\sigma|r)P(r)}{P(\sigma)}
\not\propto \sum_{r\in D} P(\sigma|r)\,.
\eeq
Note that Eq.~(\ref{noneq}) differs from that of 
\cite{Deutsch:95}, where equivalence is assumed.

\subsection{The Multisequence Method}
\label{multimethod}

A MC-based method for optimization of $P(r_0|\sigma)$ at general
$T$ has been proposed by Seno~\etal~\cite{Seno:96}. Their approach is   
based on simulated annealing in $\sigma$ with a chain-growth MC in $r$ for 
each $\sigma$. This gives a nested MC which is prohibitively time-consuming
except for very small systems.

\begin{figure}[tbh]
\vspace{-55mm}
\mbox{
  \hspace{-10mm}
  \psfig{figure=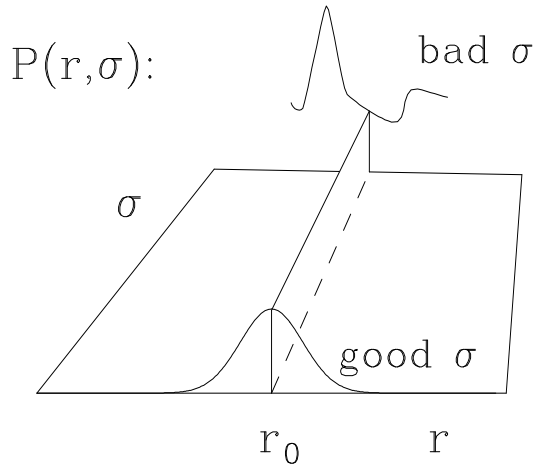,width=13cm,height=18.4cm}
}
\vspace{-65mm}
\caption{The distribution $P(r,\sigma)$. The choice of $g(\sigma)$ in 
Eq.~(\ref{g}) 
makes $P(r_0,\sigma)$ flat in $\sigma$. A sequence not designing $r_0$ 
will have maxima in $P(r_i | \sigma)$ for $r_i \not= r_0$ due to states 
with $E(r_i,\sigma)\leq E(r_0,\sigma)$. A sequence designing $r_0$ will have 
a unique maximum at $r=r_0$ in $P(r | \sigma)$, which for low $T$ 
contains most of the probability.}
\label{fig:1}
\end{figure}

The multisequence method offers a fundamentally different approach. 
In this method one replaces the simulations of $P(r|\sigma)$ for a number 
of different fixed $\sigma$ by a single simulation of the joint 
probability distribution 
\bea
P(r,\sigma) & = &\frac{1}{Z} \exp[-g(\sigma)-E(r,\sigma)/T],\\
Z & = & \sum_\sigma \exp[-g(\sigma)] Z(\sigma).  
\label{joint2}
\eea
The parameters $g(\sigma)$ determine the marginal distribution  
\beq
P(\sigma)=\frac{1}{Z}\exp[-g(\sigma)]Z(\sigma)
\label{marginal2}
\eeq
and must therefore be chosen carefully. At first
sight, it may seem that one would need to estimate $Z(\sigma)$ in order
to obtain reasonable $g(\sigma)$. However, a convenient choice is  
\beq
g(\sigma)=-E(r_0,\sigma)/T,
\label{g}
\eeq
for which one has
\beq
P(r_0|\sigma)=\frac{P(r_0,\sigma)}{P(\sigma)}=\frac{1}{ZP(\sigma)}.
\label{bayes}
\eeq
In other words, maximizing $P(r_0|\sigma)$ is in this case equivalent to 
minimizing $P(\sigma)$. This implies that bad sequences are visited 
more frequently than good ones in the simulation. This property 
may seem unattractive at a first glance. However, it can be used 
to eliminate bad sequences. The situation is illustrated in 
Fig.~\ref{fig:1}.  

The idea of using the multisequence method for sequence design
is natural since the task is to compare different sequences.  
Let us therefore stress that the method is not only convenient, 
but also efficient. The basic reason for this is that the system
often can move more efficiently through conformation space if the 
sequence degrees of freedom are allowed to fluctuate. As a result, 
simulating many sequences with the multisequence method can be 
faster than simulating a {\it single} sequence with standard 
methods, as will be shown in Sec.~\ref{sec_mult}. Another appealing 
feature of the multisequence scheme is that the optimization of the 
desired quantity $P(r_0|\sigma)$, which refers to a single structure, 
can be replaced by an optimization of the marginal probability 
$P(\sigma)$.                

The basic idea of the multisequence method is the same as in the
method of simulated tempering~\cite{Lyubartsev:92,Marinari:92,Irback:95b}. 
The only difference is that in the latter it is the temperature 
rather than the sequence which is dynamical. It has been shown 
that simulated tempering is a very efficient method for fixed-sequence 
simulations in the HP model~\cite{Irback:98b}. In particular, it was
applied to a $N=64$ sequence with known ground state, for which other
methods had failed to reach the ground state level. Simulated tempering
was, by contrast, able to find the ground state. Below we use simulated 
tempering to check our sequence design results for long chains.

\subsection{Reducing the Sequence Set}

The simple scheme outlined above is normally of little use on its own.
With a large number of sequences, it becomes impracticable, especially 
since bad sequences tend to dominate in the simulation. 
It is therefore essential to incorporate a procedure for removal  
of bad sequences. This elimination can be done in different ways. 
We will discuss two possibilities which will be referred to as 
$P(\sigma)$- and $E(r,\sigma)$-based elimination, respectively. 
Whereas both options are available for lattice models, 
$P(\sigma)$-based elimination is more appropriate for off-lattice 
models.   
  
\subsubsection{$P(\sigma)$-Based Elimination}

$P(\sigma)$-based elimination relies on the fact that bad sequences
have high $P(\sigma)$ [see Eq.~(\ref{bayes})]. The full design procedure 
consists in this case of a number of ordinary multisequence runs. After 
each of these runs $P(\sigma)$ is estimated for all the $N_r$ remaining 
sequences, and those having 
\beq	
P(\sigma)>\Lambda/N_r
\label{lambda}
\eeq
are removed. Typical values of the parameter $\Lambda$ are 1--2.   

\subsubsection{$E(r,\sigma)$-based Elimination}

The procedure referred to as $E(r,\sigma)$-based removes sequences that do not 
have the target structure $r_0$ as their unique ground state. For each
conformation $r\ne r_0$ visited in the simulation, it is checked, for 
each remaining sequence $\sigma$, whether 
\beq
E(r,\sigma)\le E(r_0,\sigma). 
\label{E_based}
\eeq
Those sequences for which Eq.~(\ref{E_based}) is true are removed. 
With this type of elimination, it may happen that one removes 
the sequence that actually maximizes $P(r_0|\sigma)$ at 
the design temperature --- the best sequence at this temperature does not 
necessarily have $r_0$ as its unique ground state (for an example, see  
Fig.~\ref{fig:2} below). This should not be 
viewed as a shortcoming of the method. If it happens, it rather means that 
the design temperature is too high. $E(r,\sigma)$-based elimination is free
from statistical errors in the sense that a sequence that does have $r_0$
as its unique ground state cannot be removed. Hence, in a very long simulation
the surviving sequences are, by construction, precisely those that have $r_0$ 
as their unique ground state.

\subsection{Restricted Search by Clamping} 
\label{clamping}

For long chains it is not feasible to explore the entire 
sequence space. On the other hand, at least in a hydrophobic/hydrophilic
model, there are typically several positions in the target structure where
$\sigma_i$ is effectively frozen  (see e.g.~\cite{Yue:95a,Li:96}).
As will be discussed below, it turns out that such positions can be 
easily detected by means of trial runs.

\section{Lattice Model Results}
\label{sec_lattice}

In this section we explore the multisequence approach on the HP 
model on the square lattice. In this context we also compare with and 
discuss other approaches; $E(r_0,\sigma)$-minimization and high-$T$ 
expansions.

The HP model contains two monomer types, H (hydrophobic) and 
P (hydrophilic/polar), and is defined by the energy 
function~\cite{Lau:89} 
\beq
\label{HP}
E(r,\sigma) = 
-\sum_{i<j}\sigma_i \sigma_j\Delta(r_i - r_j)\,,
\eeq
where $\Delta(r_i - r_j)=1$ if monomers $i$ and $j$ are 
non-bonded nearest neighbors and $0$ otherwise. For hydrophobic 
and polar monomers, one has $\sigma_i=1$ and 0, respectively.

Our explorations naturally divide into two categories; $N=16$ and 
18, where finding suitable structures and verifying folding 
properties of the designed sequences is trivial, and  
$N=32$ and 50, where this is not the case.   

For $N \leq 18$ the HP model can be solved exactly by enumeration. 
Hence such systems have been extensively used for gauging 
algorithm performances. In Table~\ref{hb} properties for $N=16$ 
and 18 systems are listed~\cite{Chan:94,Irback:98a}. A structure is 
designable if there exists a sequence for which it represents a 
unique ground state. The fraction of 
designable structures drops sharply with $N$. Furthermore, it depends 
strongly upon local interactions \cite{Irback:98a}.
\begin{table}[hbt]
\begin{center}
\begin{tabular}{|l||r|r|} 
\hline
                             & $N=16$  & $N=18$   \\
\hline \hline
\# of sequences ($2^N$)                           & 65 536   & 262 144 \\
\# of sequences with unique ground state          & 1 539    &   6 349 \\
\# of structures                                  & 802 075  & 5 808 335 \\
\# of designable structures                       &  456    &   1 475 \\
\hline
\end{tabular}
\caption{Sequence and structure statistics for the HP model for 
$N=16$ and 18.} 
\label{hb}
\end{center}
\end{table}

For a given target structure $r_0$, it is convenient to classify the 
sequences  as ``good'', ``medium'' or ``bad''. {\it Good} sequences have 
$r_0$ as their unique ground state, whereas {\it medium} sequences 
have $g>1$ degenerate ground states, one of them being $r_0$. 
Finally, {\it bad} sequences do not have $r_0$ as   
minimum energy structure. 

In our MC calculations, the elementary moves in $r$ space are standard.
Three types are used: one-bead, two-bead and pivot~\cite{Sokal:95}.     
Throughout the paper, a MC sweep refers to a combination of $N-1$ 
one-bead steps, $N-2$ two-bead steps and one pivot step. 
The new feature is that the $r$ moves are combined with stochastic 
moves in $\sigma$. Each sweep is followed by one $\sigma$ update.
The $\sigma$ update is an ordinary Metropolis 
step~\cite{Metropolis:53}. 

\subsection{N=16/18}
\label{16/18}

We have performed design calculations for a large number of
different $N=16$ and 18 target structures. Our results show that 
the multisequence design method is able to reproduce the exact data 
very rapidly. Some examples illustrating this were reported 
in~\cite{Irback:97}. 
 
Our calculations for $N=16$ and 18 are carried out using 
$E(r,\sigma)$-based elimination. Those sequences that survive the 
elimination are compared by determining their relative 
weights $P(\sigma)$, see Eq.~(\ref{bayes}). The stochastic 
$\sigma$ moves are essential in the second part of these 
calculations, when estimating $P(\sigma)$, but the first part, 
the elimination, could in principle be done without using these moves. 
In~\cite{Irback:97} it was shown, however, that it is advantageous 
to include the stochastic $\sigma$ moves in the first part 
as well. The efficiency is higher and less $T$ dependent when 
these moves are included. 
 
To make sure that the success reported in~\cite{Irback:97} was
not accidental, we applied our method to all the 1475 designable 
$N=18$ structures. For each structure we performed five experiments,  
for different random number seeds, each started from all 
$2^N$ possible sequences. Since the elimination is 
$E(r,\sigma)$-based, only the good sequences survive if the 
simulation is sufficiently long. 
The average number of MC sweeps needed to single out the 
good sequences was 123000 
(30 CPU seconds on DEC Alpha 200). A very few experiments required up to 
$10^{7}$ MC sweeps, while all five experiments converged in less 
than 500000 MC sweeps for 90\% of the structures. This shows that 
the elimination procedure is both fast and robust.

\subsection{Other Methods} 
\label{p_vs_e}

\subsubsection{Minimizing $E(r_0,\sigma)$}

Maximizing $P(r_0|\sigma)$ [see Eq.~(\ref{P})] 
is equivalent to minimizing the quantity
\beq
\Delta F_0(\sigma) = -T\ln P(r_0|\sigma)= E(r_0,\sigma)-F(\sigma)\,, 
\label{DF}
\eeq
where $F(\sigma)$ is the free energy of sequence $\sigma$
at temperature $T$. In the energy minimization method 
\cite{Shakhnovich:93b}, one approximates $\Delta F_0(\sigma)$ 
by replacing $F(\sigma)$ with a constraint that conserves the net 
hydrophobicity to a preset value $N_H$,
\beq
\label{N_H}
\sum_i \sigma_i=N_H\,.
\eeq
The reason for imposing this constraint is more 
fundamental than just guiding the sequence optimization to an 
appropriate  net hydrophobicity. In e.g. the HP model one 
has a pure ``ferromagnetic'' system in terms of $\sigma_i$ for a 
fixed $r_0$. Hence, minimizing $E(r_0,\sigma)$ with respect to 
$\sigma$ would result in a homopolymer with all monomers being 
hydrophobic. With the constraint in Eq.~(\ref{N_H}) present, this 
is avoided.

In \cite{Shakhnovich:93b} the relevant $N_H$ is picked for the 
structure to be designed.  However, this does not correspond to 
a ``real-world'' situation, where 
$N_H$ is not known beforehand. When comparing algorithm performances in 
\cite{Deutsch:96,Seno:96} a default constraint, $N_H=N/2$, was 
therefore used. Below, we will in our comparisons scan through all 
$N_H$ and minimize $E(r_0,\sigma)$ separately for each $N_H$. 

For $N=16$ and 18 all 456 respective 1475 different {\it designable} 
structures (see Table~\ref{hb}) are subject to design by minimizing  
$E(r_0,\sigma)$ for all $N_H$. If the resulting minima are 
non-degenerate for fixed $N_H$, the sequences are kept as 
candidates for good sequences, otherwise they are discarded.
A check of the results obtained this way against exact
data shows that there is at least one good sequence 
among the candidates for 87{\%}/78{\%} of the structures for $N=16$ 
and 18, respectively. In these cases we say that the method successfully
can design the structure. Another measure of the success 
of the method is given by the probability that an arbitrary 
generated candidate is good. In total, we obtained 939/3546 
candidates out of which 46{\%}/36{\%} (435/1245 sequences) are good. 
Therefore, in order to get the relatively high success rates mentioned 
above, it is essential to be able to distinguish good candidates 
from bad ones. The cost of doing this is for long chains much larger 
than that of the energy minimization itself.    

In Table~\ref{comp} the performance of the 
$E(r_0,\sigma)$-minimization methods for $N=16$ and 18 is compared 
with other approaches with respect to design ability and CPU consumption. 
As can be seen, the multisequence method with its 100\% performance, 
is indeed very fast. Furthermore, the performance of
the $E(r_0,\sigma)$-minimization variants deteriorates with size. 

\begin{table}[htb]
\begin{center}
\begin{tabular}{|l||r|r|r|r|r|} 
\cline{2-6}
\multicolumn{1}{l||}{}
& \multicolumn{2}{c|}{$E(r_0,\sigma)$-minimization} & & & \\
\multicolumn{1}{l||}{}
& $N_H=N/2$ & All $N_H$ & High-$T$ & NMC & {\bf MS} \\
\hline \hline
HP $N=16$& 25\%  & 87\% & 70\%  & 100\% & 100\%\\
HP $N=18$& 21\%  & 78\% & 50\%  & 100\% & 100\%\\
\hline
CPU sec/structure & $O$(0.1) & $O$(1) & $O$(0.1) & $O$(10$^3$) & $O$(10)\\
\hline
\end{tabular}
\caption{Number of structures that get designed by the different 
approaches for $N=16$ and 18; $E(r_0,\sigma)$-minimization 
with fixed $N_H=N/2$ and with scanning through all $N_H$, 
respectively, the nested MC approach of \protect\cite{Seno:96} (NMC), 
and the multisequence method (MS). Also shown is the computational 
demand for $N=18$ (DEC Alpha 200).}
\label{comp}
\end{center}
\end{table}

\subsubsection{High-$T$ Expansion -- Crossings}
\label{p_cross}

A more systematic approach, based on cumulant approximations 
of $F(\sigma)$, has been advocated by Deutsch and Kurosky~\cite{Deutsch:96},  
and a method along these lines has also been proposed by Morrissey 
Shakhnovich~\cite{Morrissey:96}. However, these are   
high-$T$ approximations, and can fail at relevant design 
temperatures, as has been pointed out by Seno~\etal~\cite{Seno:96}.  

The importance of the choice of the design temperature is easily 
studied for short HP chains, for which the $T$ dependence of 
$P(r_0|\sigma)$ can be computed exactly. 
At $T=0$ the relative population of $r_0$, 
$P(r_0|\sigma)$, is equal to $1$, $1/g$, and $0$ for good, 
medium, and bad sequences, respectively. 
For good sequences, the temperature at which 
$P(r_0|\sigma)=1/2$ is often referred to as the folding temperature.     

\begin{figure}[t]
\vspace{-57mm}
\mbox{
  \hspace{0mm}
  \psfig{figure=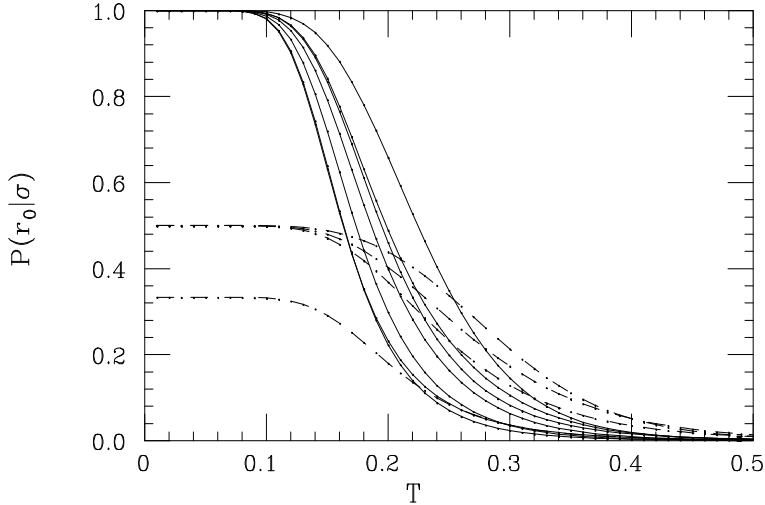,width=13cm,height=18.4cm}
}
\vspace{-57mm}
\caption{$P(r_0|\sigma)$ versus $T$ for the seven good sequences (solid) and 
for four of the crossing sequences (dashed) for the $N=18$ structure 
in~\cite{Irback:97}.}
\label{fig:2}
\end{figure}

We calculated the $T$ dependence of $P(r_0|\sigma)$ for one 
$N=16$ structure from \cite{Seno:96}, which has 1 good and 1322 
medium sequences, and one $N=18$ structure from \cite{Irback:97} 
with 7 good and 2372 medium sequences.
In the $N=18$ case, it turns 
out that there are 667 medium  sequences that have higher 
$P(r_0|\sigma)$ than at least one of the good sequences at some $T$. 
We denote these as {\it crossing} sequences. Figure~\ref{fig:2} shows 
the results for the 7 good sequences and 4 of the crossing sequences. 
In particular, one sees that in order for $P(r_0|\sigma)$ 
optimization to actually lead to a good sequence, it is necessary to 
work at a design temperature not much higher than the highest folding 
temperature. At such low temperatures, it is 
clear that high-$T$ approximations are inappropriate. 
For the $N=16$ structure it was demonstrated in \cite{Seno:96} that 
the method of \cite{Deutsch:96} fails. Indeed, it turns out that 
this structure has 296 crossing sequences. 

With these crossing phenomena, it is not surprising that the 
high-$T$ expansion frequently fails as can be seen from the summary in  
Table~\ref{comp}, from which it is also clear that the performance 
deteriorates when increasing $N$ from 16 to 18.

MC methods have the advantage that the design temperature
can be taken low enough to avoid crossing problems, without 
introducing any systematic bias. Still, in practise, it is of course not 
possible to work at too low design temperatures, due to long decorrelation  
times at low $T$. It is therefore important to note that the
multisequence 
$E(r,\sigma)$-based elimination multisequence method can be carried out at 
any temperature without running the risk of eliminating any good sequences.  

\subsection{N=32}
\label{32}

\begin{figure}[t]
\vspace{-40mm}
\mbox{
  \psfig{figure=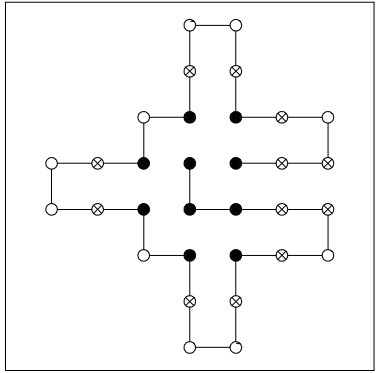,width=10.5cm,height=14cm}
}
\vspace{-43mm}
\caption{Target structure for $N=32$. Symbols are explained in the  
text.}
\label{fig:3}
\end{figure}

Having compared different methods for short chains, we now
turn to longer chains focusing on multisequence design.
For long chains it is not feasible to explore the entire 
sequence space. On the other hand, it is expected that, for a given
target structure, there are several positions along the chain where
most of the good sequences share the same $\sigma_i$ value 
(see e.g.~\cite{Yue:95a,Li:96}); in other words, some positions
are effectively frozen to H or P. A natural approach therefore is to
restrict the search by identifying and subsequently clamping 
such $\sigma_i$ to H or P. For this purpose it is convenient 
to use a set of short trial runs, as was shown in~\cite{Irback:97},
using the target structure in Fig.~\ref{fig:3}. For this structure    
ten $\sigma_i$ were clamped to H (filled circles in Fig.~\ref{fig:3})
and ten to P (open circles). Sequence optimization is then performed  
with the remaining twelve $\sigma_i$ (crosses) left open.

This clamping method can of course be generalized to 
a corresponding multi-step procedure for very long chains.

\begin{figure}[tbp]
\begin{center}
\vspace{-43mm}
\mbox{
  \psfig{figure=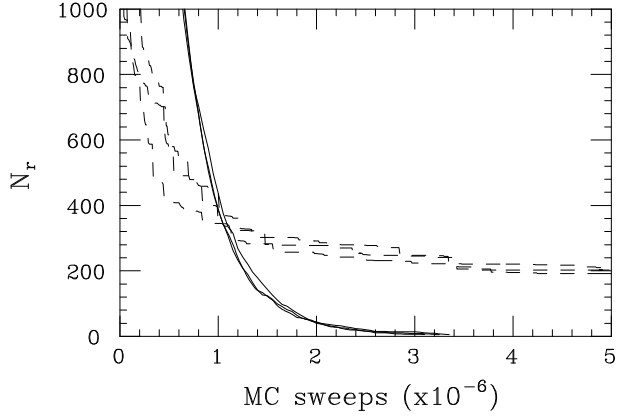,width=10.5cm,height=14cm}
}
\vspace{-43mm}
\end{center}
\caption{$N_r$ against MC time
for three runs with $P(\sigma)$-based elimination (full lines) and three with
$E(r,\sigma)$-based elimination (dashed lines) for the target structure in   
Fig.~\protect\ref{fig:3} (see text). 
}    
\label{fig:4}
\end{figure}

Taking the target structure in Fig.~\ref{fig:3} as example, with
the search restricted to $2^{12}$ sequences as described above,   
we now discuss two other important issues. First, we 
compare the efficiency of $E(r,\sigma)$-based elimination to that of 
$P(\sigma)$-based elimination.
In Fig.~\ref{fig:4} we show the number of remaining sequences, $N_r$, 
against MC time in three runs for each of the two methods 
($T=1/3$, 1 CPU hour or less per run). 
$E(r,\sigma)$-based elimination is very fast in the beginning, 
and a level is quickly reached at which it is easy to perform 
a final multisequence simulation for the remaining sequences. 
The curves level off at relatively high $N_r$, 
indicating that there are many good 
sequences for this structure (these runs were continued until all three 
contained the same 167 sequences). The three runs with $P(\sigma)$-based 
elimination, which were carried out using 50000 MC sweeps for each elimination 
step and $\Lambda=2$ [see Eq.~(\ref{lambda})], were continued until 
five sequences or fewer were left. The results were checked against 
those of the long multisequence simulations discussed in 
Sec.~\ref{sec_mult}, and were found to be quite stable in spite
of the fact that the runs were short. In particular, the best 
sequence (sequence A of Table~\ref{tab:3} below) was among the survivors 
in all three cases.    

\begin{figure}[t]
\vspace{-40mm}
\mbox{
  \hspace{-20mm}
  \psfig{figure=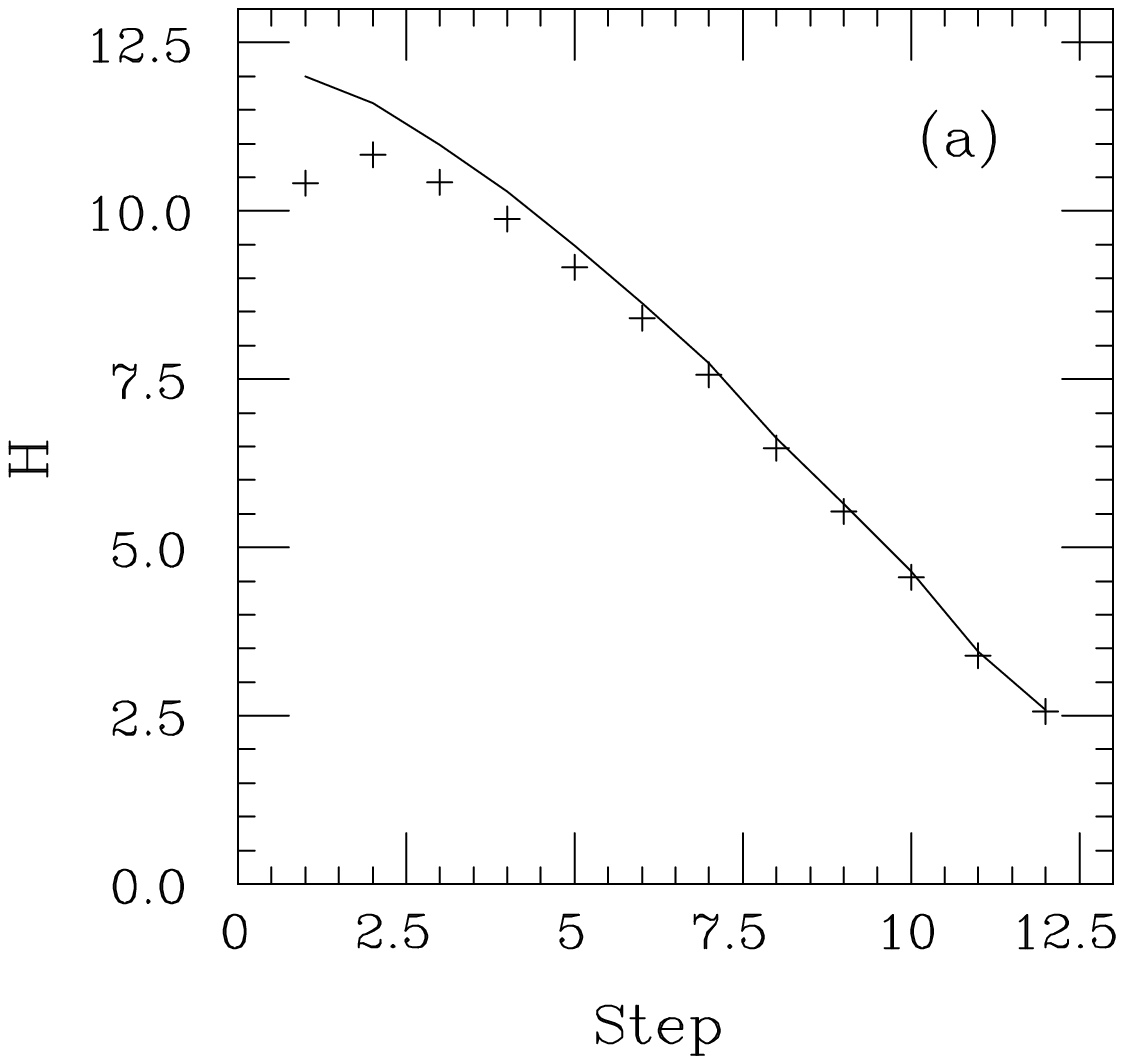,width=9cm,height=12.7cm}
  \hspace{-20mm}
  \psfig{figure=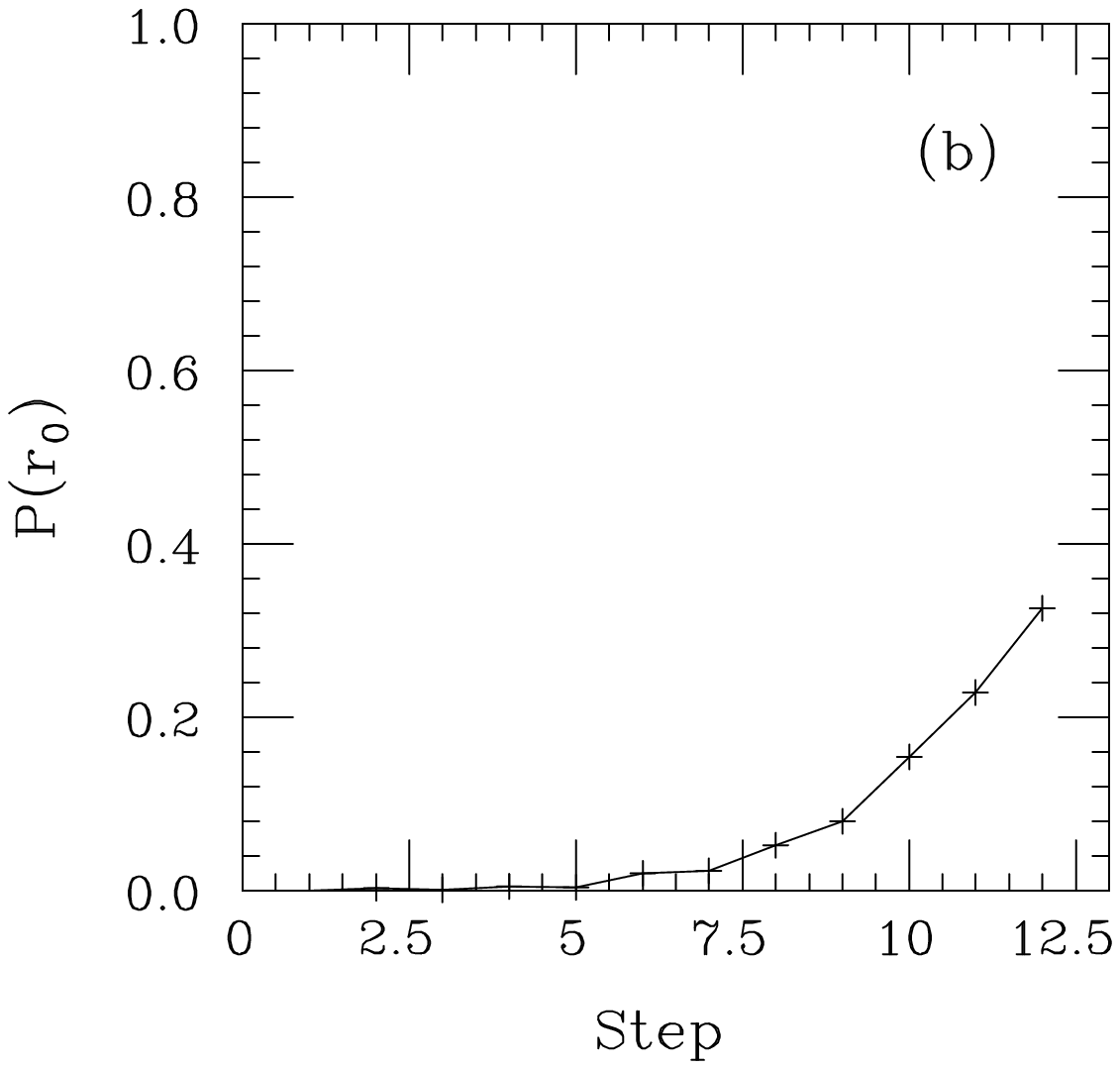,width=9cm,height=12.7cm}
}
\vspace{-35mm}
\caption{The evolution of {\bf (a)} the entropy of $P(\sigma)$ and {\bf (b)} 
the marginal probability $P(r_0)$ in a run with $P(\sigma)$-based elimination
($T=1/3$, $\Lambda=1$, $10^7$ MC sweeps for each elimination step) for the 
target structure in Fig.~\protect\ref{fig:3}. The line in (a) shows 
$\log_2 N_r$, where $N_r$ is the number of remaining sequences.} 
\label{fig:5}
\end{figure}

Next we take a look at the distribution $P(\sigma)$. 
The performance of the design procedure is crucially dependent on 
the shape of this distribution, especially when  
$P(\sigma)$-based elimination is used. One runs into problems 
if the distribution is dominated by a relatively small number of sequences 
with high $P(\sigma)$. It is therefore interesting to see how the shape of 
$P(\sigma)$ evolves as the elimination process proceeds. Figure~\ref{fig:5}a 
shows the entropy of $P(\sigma)$,
\beq
H=-\sum_\sigma P(\sigma)\log_2P(\sigma),
\label{entropy}
\eeq 
in a run with $P(\sigma)$-based elimination. With $N_r$
remaining sequences, the maximal value of $H$ is $\log_2 N_r$, corresponding 
to a uniform distribution $P(\sigma)$. As can be seen from
Fig.~\ref{fig:5}a,  
after a few elimination steps, $H$ is close to this limit. The desired
behavior of the marginal distribution of $r$, $P(r)$, is in a sense the 
opposite, since the weight of the target structure should become large. 
The evolution of $P(r_0)$ in the same run is shown in Fig.~\ref{fig:5}b.

\subsection{N=50}
\label{50}

A test of any design procedure consists of three steps:

{\bf 1.} Finding a suitable target structure.\\ 
{\bf 2.} Performing the actual design.\\
{\bf 3.} Verifying that the final sequence is good. 

In this section we discuss the design of a $N=50$ 
structure. For this system size the first step is highly 
non-trivial. Also, the verification part is quite  
time-consuming. For these reasons we focus on one example 
and go through each of the steps in some detail.    

\subsubsection{Finding a Suitable Target Structure}

We begin with the problem of finding a suitable target structure.
For a randomly chosen structure it is unlikely that there is any 
sequence that can design it; the fraction of designable structures 
is e.g. about 0.00025 for $N=18$ (see Table~\ref{comp}). Furthermore, 
this fraction decreases with system size.  
Rather than proceeding by trial and error, we therefore 
determined the target structure by employing a variant of our 
sequence design algorithm. In this version no target structure  
is specified and Eq.~(\ref{g}) is replaced by      
\beq
g(\sigma)=-E_{\min}(\sigma)/T, 
\label{no_fix}
\eeq
where $E_{\min}(\sigma)$ ideally should be the minimum energy for the
sequence $\sigma$. In our calculations, since the minimum energy is 
unknown, we set $E_{\min}(\sigma)$ equal to the lowest energy 
encountered so far. Except for this change of the parameters $g(\sigma)$, 
we proceed exactly as before, using $P(\sigma)$-based elimination. 
However, a sequence is never eliminated if its $g(\sigma)$ was 
changed during the last multisequence run, that is if a new lowest 
energy was found.

With this algorithm, one may hope to identify and eliminate those sequences 
that are bad not only with respect to one particular structure, but with 
respect to all possible structures. Clearly, this is a much more ambitious 
goal, and it should be stressed a careful evaluation of the usefulness of this 
approach is beyond the scope of the present paper. 

\begin{figure}[t]
\vspace{-41mm}
\mbox{
  \hspace{-40mm}
  \psfig{figure=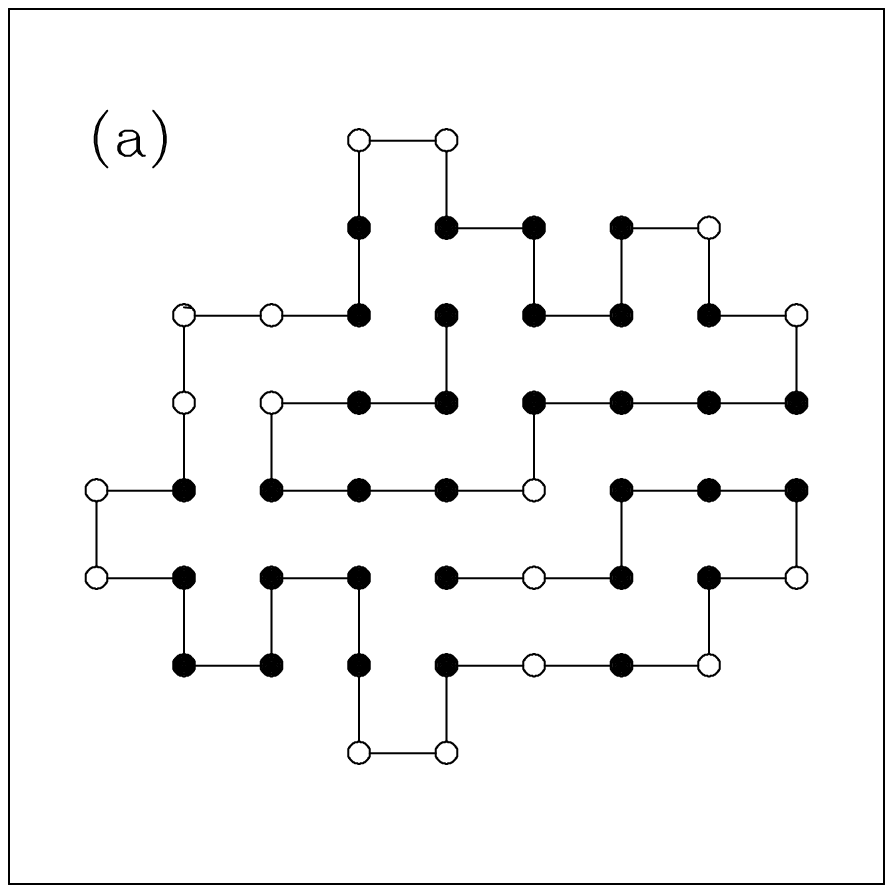,width=10.5cm,height=14cm}
  \hspace{-30mm}
  \psfig{figure=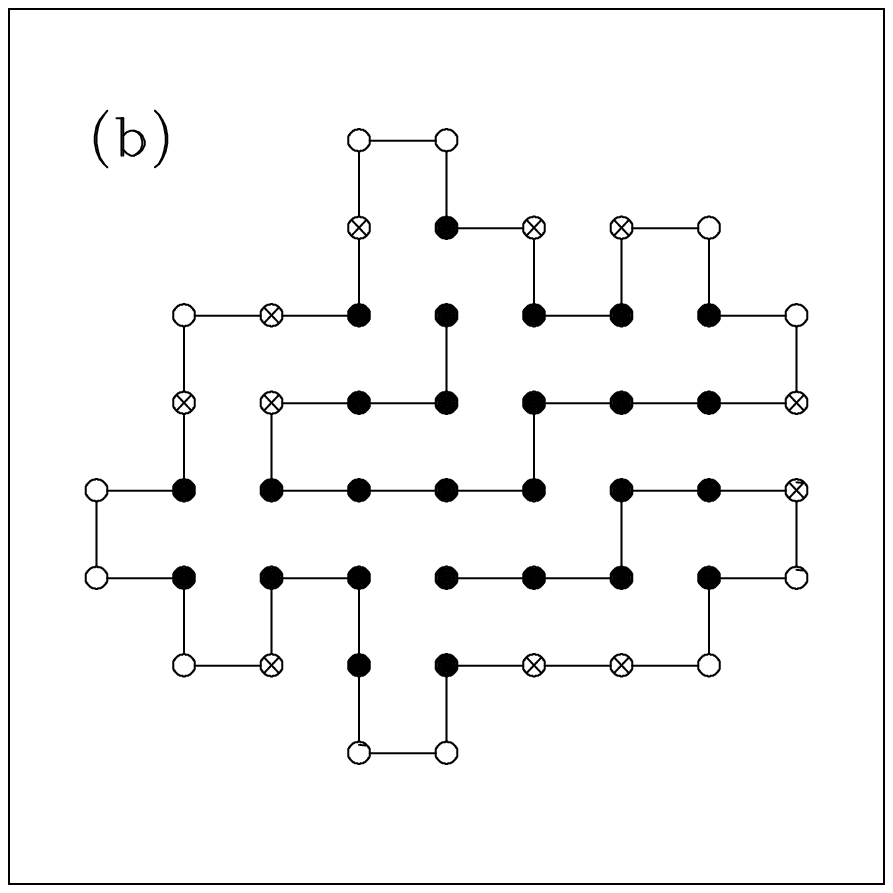,width=10.5cm,height=14cm}
}
\vspace{-44mm}
\caption{{\bf (a)} Target structure for $N=50$. This structure and 
the sequence shown were obtained using the design algorithm without
fixed target [see Eq. (\protect\ref{no_fix})]. {\bf (b)} Results of 
the clamping procedure for our $N=50$ target structure. Symbols are 
the same as in Fig.~\protect\ref{fig:3}.}  
\label{fig:6}
\end{figure}

This calculation was started from a set of about 2200 sequences. 
These were obtained by first randomly generating a mother sequence, with 
probability 0.65 for H, and then randomly changing this at one 
to three positions. Thus, there is a high degree of similarity 
between the sequences, which ensures a reasonable 
acceptance rate for the sequence update. After 37 elimination steps 
($T=1/2.8$, $\Lambda=1.5$, $2\times10^5$ MC sweeps for each elimination step), 
three of the sequences were left. The best of these sequences
and its minimum energy structure can be found in Fig.~\ref{fig:6}a. 
Note that this sequence does not minimize the energy for 
any fixed $N_H$ --- the energy can be reduced by interchanging 
the monomers $i=19$ and 43 ($i=1$ corresponds to the lowest of the 
two end points in Fig.~\ref{fig:6}a). In what 
follows we take this structure as our target structure, without using any 
information about the particular sequence shown.

\subsubsection{Sequence Design}

We began the sequence design for this structure by performing 
ten short runs, each started from $10^5$ random sequences. 
Based on these, 27 $\sigma_i$ were clamped to H and 12 to P,
as illustrated in Fig.~\ref{fig:6}b. It is interesting to 
compare these results to the original sequence 
in Fig.~\ref{fig:6}a. As expected, there is a close similarity, 
but there are also three positions along the chain at which 
$\sigma_i$ is clamped to the opposite value compared to the 
original sequence ($i=2$, 19 and 43). Thus, the original sequence does 
not belong to the restricted sequence set which we study next. 

Having restricted the search,
we proceed in two steps. First, we apply $E(r,\sigma)$-based 
elimination. As in the corresponding $N=32$ calculation, 
the number of remaining sequences rapidly reached a fairly stable 
and high level, indicating that there are many sequences with the 
target structure as unique ground state. The number of sequences surviving 
this first step was 832. The second step is a simulation 
with $P(\sigma)$-based elimination 
($T=1/2.8$, $\Lambda=1$, $10^7$ MC sweeps for each elimination step).
This step was repeated three times using different random number 
seeds, each time starting from the same 832 sequences. The stability
of the results was not perfect, but the best sequence found was the
same in all three runs. This sequence has four H and seven P at the
positions left open after clamping. The four positions 
that were assigned an H are $i=10$, 11, 18 and 28.   

\subsubsection{Verification}
\label{verification}
    
In order to check the designed sequence, we performed an independent 
simulated-tempering calculation. As mentioned in 
Sec.~\ref{multimethod}, in a previous study~\cite{Irback:98b},
this method successfully found the ground state of a $N=64$ HP 
chain~\cite{Irback:98b}. The length of our $N=50$ simulation is 
$2\times 10^9$ MC sweeps.
In the simulation of the designed $N=50$ sequence the target
structure was visited many times; we estimate the number of 
``independent'' visits to be about 30. By contrast,
no other structure with the same or lower energy was encountered.   
We take this as strong evidence that the target structure 
indeed is a unique energy minimum for this sequence.    

Similar simulations were also performed for two other $N=50$
sequences, S1 and S2. The sequence S1 is the one 
shown in Fig.~\ref{fig:6}a, and S2 is the one obtained by 
assigning P to all open positions in Fig.~\ref{fig:6}b.
At first sight S1 may not seem to fit the target structure
very well; as already noticed, this sequence does
not minimize the energy of the target structure for fixed $N_H$.   
Nevertheless, our results suggest that both S1 and S2, like
the designed sequence, have the target structure as unique ground state. 
However, the dominance of this structure sets in at a lower temperature 
for S1 and S2 than for the designed sequence; rough estimates of 
the folding temperatures are 0.27 for the designed sequence
and 0.23 for S1 and S2.   

Unfortunately, it was not feasible to evaluate alternative
methods for this system size, because the verification part
is too time-consuming. Let us note, however, that our designed
sequence uniquely minimizes the energy of the target structure  
for fixed $N_H=31$. Sequence S1, on the other hand, appears to
be good too, even though it does not minimize the target energy 
for any $N_H$.

\subsection{The Folding Transition for N=50}

\begin{figure}[t]
\vspace{-40mm}
\mbox{
  \hspace{-30mm}
  \psfig{figure=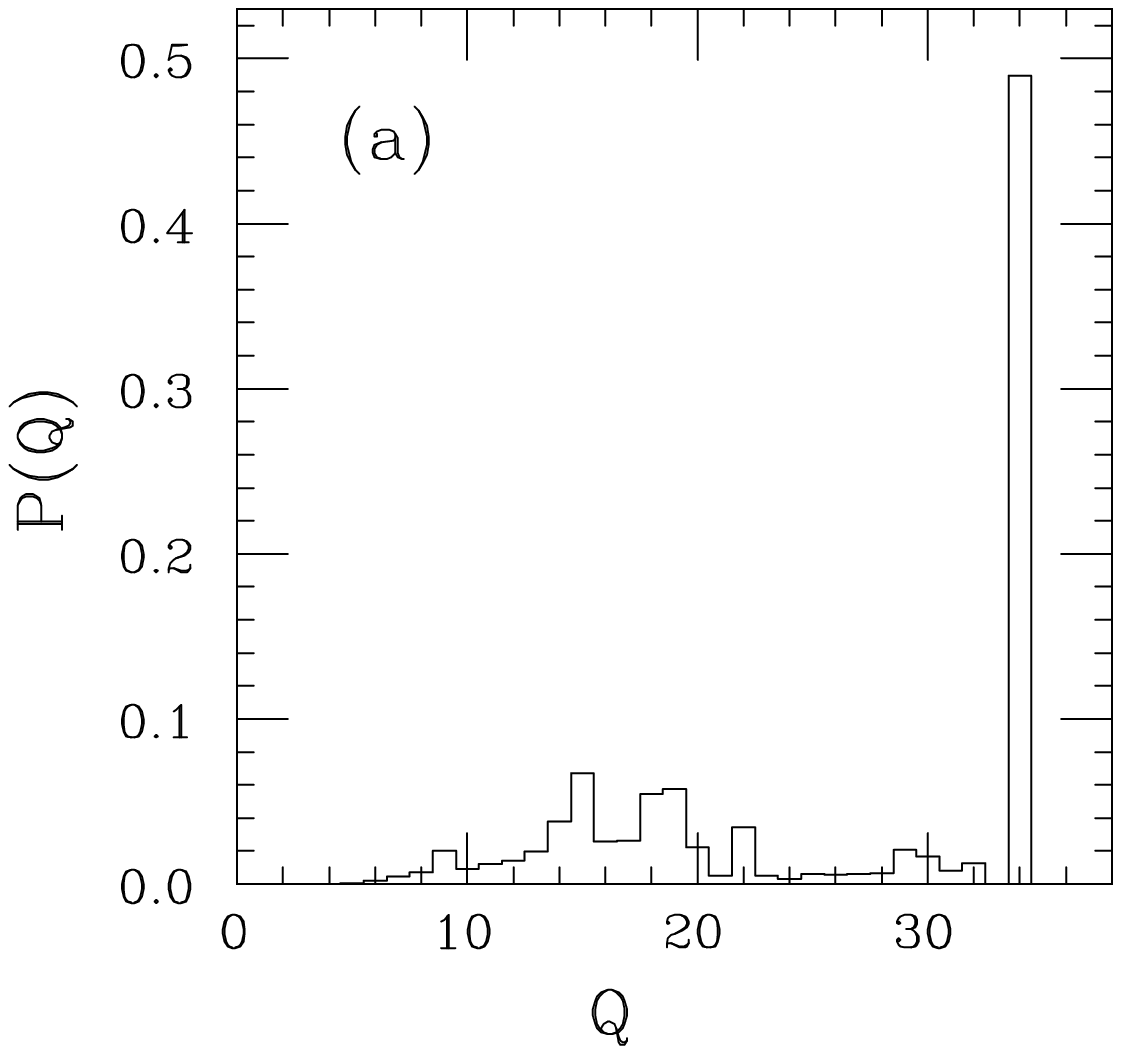,width=10.5cm,height=14cm}
  \hspace{-30mm}
  \psfig{figure=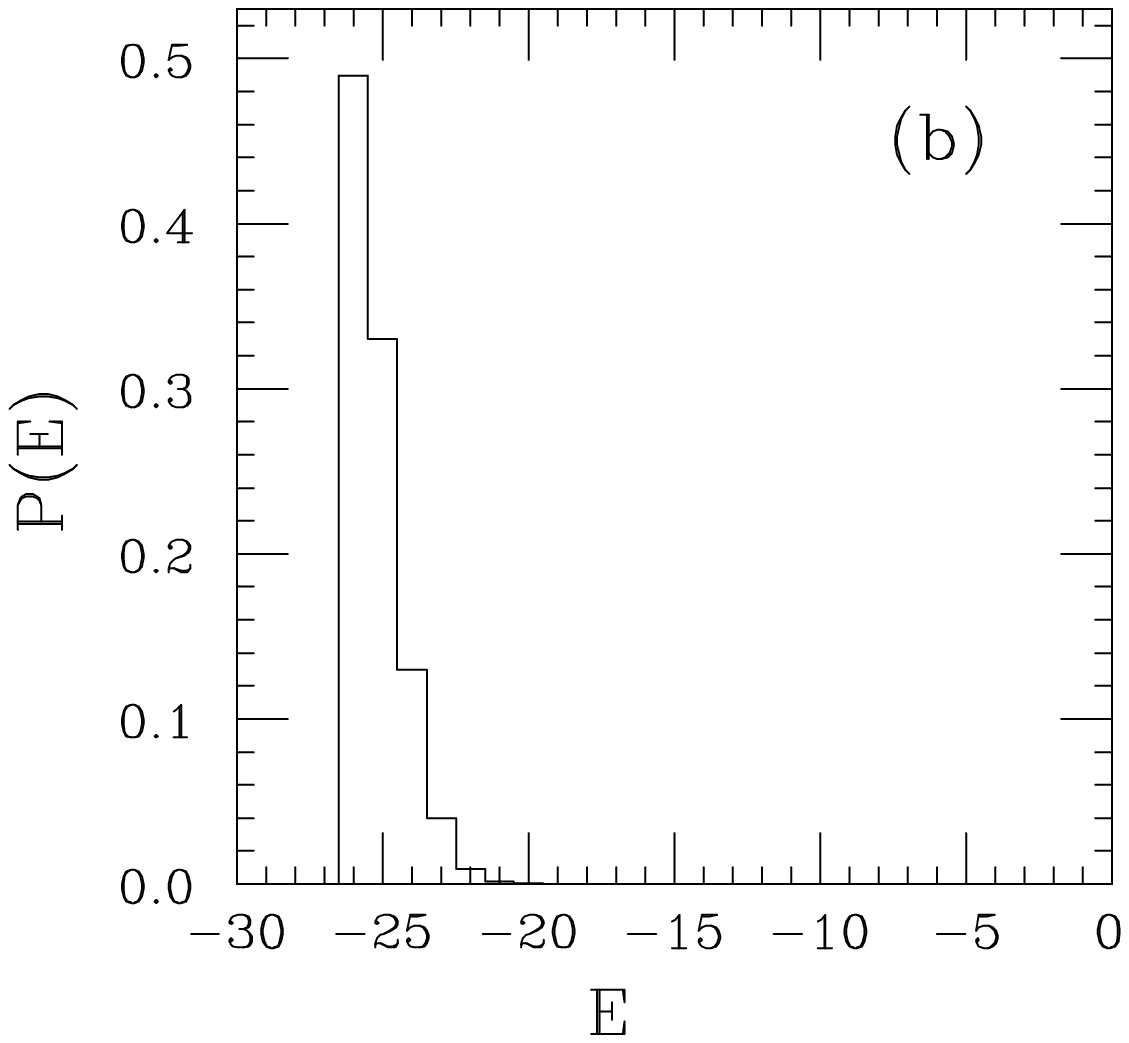,width=10.5cm,height=14cm}
}
\vspace{-43mm}
\caption{The probability distributions of {\bf (a)}
the similarity parameter $Q$ and {\bf (b)} the energy $E$ for 
the $N=50$ sequences S2 at $T=0.227\approx T_f$.}
\label{fig:7}
\end{figure}

The simulated-tempering runs for the three $N=50$ sequences 
provide thermodynamic data over a wide range of temperatures.
In particular, they offer an accurate picture of the behavior 
at the folding transition. Shown in Fig.~\ref{fig:7} are the 
probability distributions of the similarity parameter $Q$ 
(number of contacts that a given conformation shares with the 
native state) and the energy $E$, close to the folding 
temperature $T_f$ for sequence S2. 
The corresponding results for the other two sequences are 
qualitatively similar.

From Fig.~\ref{fig:7}a it can be seen that the distribution $P(Q)$
has an essentially bimodal shape. The peak at $Q=Q_{\max}=34$ 
corresponds to the native state and contains, by definition, around 50\% 
of the distribution. The non-native peak, at $Q/Q_{\max}\approx0.4-0.6$,
is well separated from the native one. 
%The separation between
%the native and non-native parts is more clear than for smaller systems
%in this model (see e.g.~\cite{Dill:95}), and 
showing that the transition is cooperative in the sense that the system
is either in the native state or in states that are structurally 
very different. It must be stressed, however, that it is not 
a two-state transition --- the non-native part does not correspond  
to one ensemble of unfolded structures, but rather to a number
of distinct folded low-energy states. The ruggedness of the 
non-native peak of $P(Q)$ is an indication of this, 
and it becomes evident from the energy distribution of  
Fig.~\ref{fig:7}b, which shows no trace of 
bimodality. The fact that it is not a two-state transition is in
line with general arguments for two-dimensional 
models~\cite{Abkevich:95,Shakhnovich:97}.  

\section{The Multisequence Method}
\label{sec_mult}

The multisequence method, which is a key ingredient in our design algorithm, 
was originally applied to a simple off-lattice model~\cite{Irback:95b} 
in the context of folding studies.
Using parameters $g(\sigma)$ 
that had been adjusted so as to have an approximately uniform 
distribution in $\sigma$, it was found to be much
more efficient than a standard MC. In this paper we have instead
chosen $g(\sigma)$ according to Eq.~(\ref{g}). This simple choice is 
not possible for a random set of sequences. The efficiency can, however,
be quite good after removal of bad sequences. To illustrate this, we 
take a set of 180 surviving $N=32$ sequences, from one of the three runs 
with $E(r,\sigma)$-based elimination in Fig.~\ref{fig:4}. 

For these sequences we carried out multisequence simulations at 
three different temperatures, $T$=1/2.8, 1/3.1 and 1/3.4. The results 
of these simulations are compared to those of single-sequence 
simulations with identical $r$ updates, for the three different sequences 
shown in Table~\ref{tab:3}. Sequence A is the best sequence 
found among all the 180. As can be seen from Table~\ref{tab:4},  
it has a folding temperature close to $T=1/3.1$. Sequences B and 
C were deliberately chosen to represent different types of behavior,
and have lower folding temperatures. 
It is interesting to note how different the sequences A and C behave
(see Table~\ref{tab:4}), in spite of the fact that they differ only by an 
interchange of two adjacent monomers.

\begin{table}
\begin{center}
\begin{tabular}{|c|c|}
\hline
Sequence A & HHPP HHPP PPHP HPPP PHPH PPPP HHPP HHHH \\
Sequence B & HHHP HHPP PPHP HHPP PHPH PPPP HHPH HHHH \\
Sequence C & HHPP HHPP PPHP HPPP PHPH PPPP HHPH PHHH \\
\hline
\end{tabular}
\caption{Three $N=32$ HP sequences.}
\label{tab:3}
\end{center}
\end{table} 

\begin{table}
\begin{center}
\begin{tabular}{|c|l|l|l|l|} 
\cline{3-5}
\multicolumn{2}{c|}{}
&\multicolumn{1}{c|}{$1/T=2.8$}
&\multicolumn{1}{c|}{$1/T=3.1$}
&\multicolumn{1}{c|}{$1/T=3.4$}\\
\hline
Sequence A&Standard MC  &$0.227\pm0.005$&$0.532\pm0.010$&$0.752\pm0.015$\\
          &Multisequence&$0.234\pm0.006$&$0.520\pm0.010$&$0.732\pm0.008$\\
\hline
Sequence B&Standard MC  &$0.0389\pm0.0024$&$0.086\pm0.007$&$0.166\pm0.021$\\
          &Multisequence&$0.0383\pm0.0013$&$0.095\pm0.003$&$0.166\pm 0.006$\\
\hline
Sequence C&Standard MC  &$0.00251\pm0.00012$&$0.0066\pm0.0003$&$0.0133\pm0.0008$\\ 
          &Multisequence&$0.00250\pm0.00009$&$0.0066\pm0.0003$&$0.0123\pm0.0005$ \\
\hline
\end{tabular}
\caption{Comparison of results for $P(r_0|\sigma)$ obtained by two different 
methods, the multisequence algorithm and a standard fixed-sequence MC. 
Shown are results for the three sequences in
Table~\protect\ref{tab:3} for three different temperatures.} 
\label{tab:4}
\end{center}
\end{table}

As the number of sweeps is the same, $10^9$, and since the cost of the 
additional sequence moves in the multisequence runs is negligible, 
we can directly compare the statistical errors from these runs. 
In Table~\ref{tab:4} the averages and statistical errors for the 
quantity $P(r_0|\sigma)$ are shown. The errors quoted are 1$\sigma$ errors, 
obtained by a jackknife procedure. 

From Table~\ref{tab:4} it can be seen that the two methods give similar
statistical errors at the highest $T$ studied, which lies above the folding 
temperature for all three sequences. It should be stressed that equal 
errors implies that the multisequence method is faster by a factor of 180, 
since a single run covers all sequences with this method. Although there is 
a dependence upon sequence, there is furthermore a clear tendency that the 
errors from the multisequence runs get smaller than those from the 
single-sequence runs at lower $T$. The difference is largest for sequence B 
and the lowest temperature. In this case the errors differ by a factor of 
3.5, which corresponds to an extra factor of 10 in computer time, in addition 
to the trivial factor of 180. 

This simple choice of $g(\sigma)$ [Eq.~(\ref{g})] has been used with 
success in all our calculations. Nevertheless, let us finally note that 
multisequence design can also be applied using other $g(\sigma)$ values. 
In particular, it is easy to modify Eq.~(\ref{bayes}) for general 
$g(\sigma)$.
  
\section{Off-Lattice Model Results}
\label{sec_off-lattice}

Lattice models offer computational and pedagogical advantages, 
but the results obtained on the lattice must be interpreted with 
care; for example, it is has been shown that the number of designable 
structures drastically depends on the lattice type in the HP 
model~\cite{Irback:98a}. In this section we therefore show that our 
design procedure can be applied essentially unchanged to a 3D 
minimalist off-lattice model~\cite{Irback:96c}. While similar 
models have been studied before, 
see e.g.~\cite{Irback:96b,Veitshans:97,Sasai:95,Nymeyer:98}, 
it is the first time, as far as we know, that sequence design is 
performed in an off-lattice model based on sampling of the full 
conformation space.       

One problem encountered in going to off-lattice models is in the
very formulation of the stability criterion. Clearly, it is the probability
of being in the vicinity of the target structure $r_0$ that we are 
interested in, rather than the probability of being precisely
in $r_0$. While this point can be relevant for lattice models
too, it is of more obvious importance in the off-lattice case. 
Throughout this paper, we stick to the probability $P(r_0|\sigma)$
corresponding to a single target structure $r_0$, using target  
structures that are obtained by energy minimization. In the general case,  
it might be necessary to consider instead the off-lattice analogue         
of the left hand side of Eq.~(\ref{noneq}). 

Another problem is the elimination criterion for bad sequences. 
A straightforward implementation of $E(r_0,\sigma)$-based elimination 
requires the introduction of a cutoff in structural similarity to 
$r_0$, below which elimination should not take place. However, 
with such a cutoff, the method 
is too slow, since  
in order to have a reasonable elimination rate, it appears necessary 
to employ some sort of quenching procedure, which tends to be very 
time-consuming. By contrast, we found $P(\sigma)$-based elimination
to be useful for off-lattice chains too, without any modifications 
or additional parameters.  

For the off-lattice model in contrast to the HP model, one does 
not have access to a set of small $N$ exact enumerations results. 
Hence, for all sizes we need to go through the three steps needed for 
$N > 18$ HP chains: find suitable structures, perform design 
and verify that the designed sequence is stable in the desired structure. 

\subsection{The Model}

Like the HP model, the 3D off-lattice model~\cite{Irback:96c} 
contains two kinds of residues, hydrophobic  ($\sigma_i=1$) and 
hydrophilic ($\sigma_i=0$). Adjacent residues are linked by rigid 
bonds of unit length, $\bv_i$, to form linear chains. The energy 
function is given by
\beq
E(\bv; \sigma) = -\kappa_1\sum_{i=1}^{N-2}\bv_i\cdot\bv_{i+1}  
- \kappa_2\sum_{i=1}^{N-3}\bv_i\cdot\bv_{i+2} + 
4\sum_{i=1}^{N-2}\sum_{j=i+2}^N 
\epsilon(\sigma_i,\sigma_j)
\left( \frac{1}{r_{ij}^{12}}-\frac{1}{r_{ij}^{6}}\right)\ 
\label{energy}
\eeq
where $r_{ij}$ denotes the distance between residues $i$ and $j$. 
The first two {\it sequence-independent} terms define the local 
interactions, which turn out to be crucial for native structure 
formation \cite{Irback:96c}. The parameters are chosen as $\kappa_1=-1$ 
and $\kappa_2=0.5$ in order to obtain thermodynamically stable 
structures, and to have local angle distributions and bond-bond correlations
that qualitatively resemble those of functional proteins.
The third term represents the {\it sequence-dependent} global 
interactions modeled by a Lennard-Jones  potential.
The depth of its minimum, $\epsilon(\sigma_i,\sigma_j)$, is chosen 
to favor the formation of a core of hydrophobic residues by setting 
$\epsilon(0,0)=1$,  $\epsilon(1,1)=\epsilon(0,1)=\epsilon(1,0)=\frac{1}{2}$.

To monitor structural stability we use the mean-square
distance $\delta_{ab}^2$ between two arbitrary configurations $a$ 
and $b$. An informative measure of stability 
is given in terms of the probability distribution $P(\delta^2)$ of 
$\delta_{ab}^2$, and the corresponding mean, 
$\langle \delta^2\rangle$.
The latter is small if the structural fluctuations are small, 
but tells nothing about the actual structure. In addition, we therefore 
measure the similarity to the desired structure $r_0$.
For this purpose we average $\delta_{ab}^2$ over configuration $a$, 
keeping configuration $b$ fixed and equal to $r_0$.  
This average will be denoted by $\ev{\delta_0^2}$.       

When investigating thermodynamic properties of 
this model one finds a strong dependence upon the local interactions. 
This impact of local interactions is {\it not} a peculiar property 
for off-lattice models. Indeed, similar findings have been   
reported for the HP lattice model~\cite{Irback:98a}.

\subsection{Design Results} 
 
\subsubsection*{Finding Suitable Structures} 

We have determined the global energy minima, or native structures, for a 
number of $N=16$ sequences, and six of these structures are used as target 
structures in our design calculations. In addition, we consider six $N=20$ 
target structures, which are native states of sequences studied 
in~\cite{Irback:96c}. We restrict ourselves to these twelve 
examples because the verification of the design results, the computation 
of $\ev{\delta_0^2}$, is time-consuming. This selection of structures 
studied represent no bias with respect to the performance of the 
design algorithm. As can be seen from Tables~\ref{s_16} and~\ref{s_20}, 
some of the original sequences represent good folders (small $\ev{\delta^2}$) 
whereas others do not (large $\langle \delta^2 \rangle$). An example 
of a $N=20$ target structure can be found in~\cite{Irback:96c} .

\subsubsection*{Designing the Sequences}

As discussed above, in our off-lattice calculations, we use
$P(\sigma)$-based elimination, which, unlike $E(r,\sigma)$-based
elimination, can be used as it stands. All our design calculations are 
carried out at the temperature $T=0.3$, whereas the highest folding
temperatures measured in \cite{Irback:96c} are close to 0.2. This
somewhat high design temperature was chosen in order to speed up the 
calculations. It is still low enough for design of stable sequences, 
as will become clear from the verification below. These verification 
calculations are performed at $T=0.15$, using simulated tempering.      
                          
Our $P(\sigma)$-based design calculations starts out from the set of all 
$2^N$ possible sequences. Each iterative step amounts to a relatively 
short multisequence simulation consisting of 500000 MC cycles for the 
$N_r$ remaining sequences, followed by removal of those sequences 
for which the estimated $P(\sigma)$ fulfills Eq.~(\ref{lambda}) with 
$\Lambda=1.5$. This is continued until a single sequence remains, which 
typically requires around 150 steps. The final sequence we take as the MS  
designed sequence. Each MC cycle consists of one attempt to update the
conformation and one for the sequence. The conformation update is either   
a rotation of a single bond $\hat{b}_i$ or a pivot move.  
The time consumption for the studied $N=16$ and 20 chains 
ranges from three to six CPU hours.

In our multisequence and simulated-tempering simulations each MC 
sweep in conformation space is followed by one attempt to update the 
sequence or temperature. The sequence and temperature updates are both 
ordinary Metropolis steps~\cite{Metropolis:53}. The CPU cost of these 
updates is negligible compared to that of the conformation update.  

The designed sequences are shown in Tables~\ref{s_16} and~\ref{s_20} for 
$N=16$ and 20, respectively. Also shown are the results
of ``naive'' energy minimization~\cite{Shakhnovich:93b}.  
Ideally, one should 
use this method by scanning through all possible $N_H$ [Eq.~(\ref{N_H})], 
which was done for $N \leq 18$ HP chains in Sec.~\ref{p_vs_e}. 
However, given that the verification is quite tedious, we have chosen 
to use a single $N_H$ only, corresponding to the original sequence. In 
other words the $E(r_0,\sigma)$-minimization method is given a slight  
advantage as compared to what would have been the case for a real-world 
application. 

\begin{table}[h]
\begin{center}
\begin{tabular}{|l|r|l|l|l|}
\cline{2-5}
\multicolumn{1}{l|}{} & method & $\sigma$ & 
$\ev{\delta^2}_{T=0.15}$ & $\ev{\delta_0^2}_{T=0.15}$\\ 
\hline
16-1  & target  &  1111100101101111 & 0.01 $\pm$ 0.0002  & 0.01       $\pm$ 0.002  \\ 
 &{\bf MS}      &  1111100101111111 & {\bf 0.01} $\pm$ 0.0002  & {\bf 0.01} $\pm$ 0.002  \\
 &$E(r_0,\sigma)$     &  1111100101011111 & 0.02 $\pm$ 0.003   & 0.01       $\pm$ 0.007  \\       
\hline
16-2  & target  &  1011001110011110 & 0.07  $\pm$ 0.003   &  0.04       $\pm$ 0.004 \\ 
 &{\bf MS}      &  1011001110011110 & {\bf 0.03}  $\pm$ 0.004   &  {\bf 0.02} $\pm$ 0.007 \\ 
 &$E(r_0,\sigma)$     &  1111001010101110 & 0.38  $\pm$ 0.03    &  0.52       $\pm$ 0.02  \\  
\hline
16-3  & target  &  1010101001101111 & 0.24  $\pm$ 0.05  &  0.13       $\pm$ 0.02  \\
 &{\bf MS}      &  1111111101001111 & {\bf 0.01}  $\pm$ 0.001 &  {\bf 0.01} $\pm$ 0.006 \\  
 &$E(r_0,\sigma)$     &  1010101101001111 & 0.08  $\pm$ 0.02  &  0.04       $\pm$ 0.02  \\ 
\hline 
16-4  & target  &  1101101000010011 & 0.38  $\pm$ 0.02 & 0.25       $\pm$ 0.02 \\ 
 &{\bf MS}      &  1111101111010011 & {\bf 0.12}  $\pm$ 0.01 & {\bf 0.36} $\pm$ 0.01 \\
 &$E(r_0,\sigma)$     &  1010101000010111 & 0.28  $\pm$ 0.01 & 0.24       $\pm$ 0.02 \\ 
\hline
16-5  & target  &  1001110011111111  & 0.47  $\pm$ 0.02  & 0.33       $\pm$ 0.02 \\
 &{\bf MS}      &  1001110010111111  & {\bf 0.12}  $\pm$ 0.002 & {\bf 0.10} $\pm$ 0.01 \\  
 &$E(r_0,\sigma)$     &  1011110010111111  & 0.11  $\pm$ 0.004 & 0.11       $\pm$ 0.01 \\ 
\hline 
16-6  & target  &  1110010000000110  & 0.64  $\pm$ 0.007 &  0.57       $\pm$ 0.02 \\
 &{\bf MS}      &  1101111110101111  & {\bf 0.30}  $\pm$ 0.02  &  {\bf 0.34} $\pm$ 0.02 \\  
 &$E(r_0,\sigma)$     &  0101010000101010  & 0.28  $\pm$ 0.02&  0.42         $\pm$ 0.01 \\ 
\hline 

\end{tabular}  
\caption{Design results for six $N=16$ off-lattice target structures. 
For each structure three sequences are listed together with the 
corresponding  $\ev{\delta^2}$ and $\ev{\delta_0^2}$: 
the sequence used to generate the target structure (``target''), and 
the sequences obtained by multisequence design ({\bf MS}) and  
$E(r_0,\sigma)$-minimization, respectively.}
\label{s_16}
\end{center}
\end{table}

\begin{table}[h]
\begin{center}
\begin{tabular}{|l|r|l|l|l|}
\cline{2-5}
\multicolumn{1}{l|}{} & method & $\sigma$ & 
$\ev{\delta^2}_{T=0.15}$ & $\ev{\delta_0^2}_{T=0.15}$\\ 
\hline
20-1  & target    &  11110011110110111001 & 0.08 $\pm$ 0.01 & 0.04 $\pm$0.01 \\ 
  & {\bf MS}      &  11110011110010111001 & {\bf 0.02} $\pm$ 0.001 & {\bf 0.01} $\pm$0.002 \\ 
  & $E(r_0,\sigma)$     &  11110011111110101001 & 0.27 $\pm$ 0.04 & 0.29 $\pm$ 0.01 \\   
\hline   
20-2  & target    &  11110110101100111011 & 0.27 $\pm$ 0.05  & 0.15$\pm$ 0.01 \\ 
  & {\bf MS}      &  11110100100100111111 & {\bf 0.02} $\pm$ 0.004 & {\bf 0.01}$\pm$ 0.003 \\
  & $E(r_0,\sigma)$     &  11110010101010111111 & 0.24 $\pm$ 0.05  & 0.93$\pm$ 0.01 \\
\hline   
20-3  & target    &  11100100101001010101 & 0.30 $\pm$ 0.04  & 0.38  $\pm$ 0.01 \\ 
  & {\bf MS}      &  11111100101001010111 & {\bf 0.10}$\pm$ 0.02  & {\bf 0.12} $\pm$ 0.01   \\
  & $E(r_0,\sigma)$     &  10101000101001010111 & 0.59 $\pm$ 0.02  & 0.53 $\pm$ 0.01  \\
\hline   
20-4  & target    &  01101111010110111110 & 0.24 $\pm$ 0.02  & 0.34$\pm$ 0.01\\ 
  & {\bf MS}      &  01101010010111111110 & {\bf 0.05} $\pm$ 0.01  & {\bf 0.03} $\pm$ 0.01 \\
  & $E(r_0,\sigma)$     &  01101011010111111110 & 0.10 $\pm$ 0.01   & 0.05$\pm$ 0.01\\
\hline 
20-5  & target    &  01111110111101101100 & 0.46 $\pm$ 0.04  & 0.29 $\pm$ 0.01 \\ 
  & {\bf MS}      &  11111110100101111101 & {\bf 0.46} $\pm$ 0.04  & {\bf 0.43}$\pm$ 0.01  \\
  & $E(r_0,\sigma)$     &  01111110100101111101 & 0.65 $\pm$ 0.09  & 0.46$\pm$ 0.01 \\
\hline 
20-6  & target    & 01100111000101011010  & 0.73 $\pm$ 0.01  & 0.75 $\pm$0.01 \\ 
  & {\bf MS}      & 11100111001101011111  & {\bf 0.64} $\pm$ 0.02  & {\bf 0.73} $\pm$ 0.01 \\
  & $E(r_0,\sigma)$     & 01111010100101001010  & 0.52 $\pm$ 0.08  & 0.91 $\pm$ 0.01 \\

\hline 
\end{tabular}
\caption{Design results for six $N=20$ off-lattice target structures. The 
corresponding sequences are those from Table 1 in~\protect\cite{Irback:96c} 
but here ordered according to decreasing $\ev{\delta^2}$.
Same notation as in Table~\ref{s_16}.}  
\label{s_20}
\end{center}
\end{table}

\subsubsection*{Verification}

To assess the quality of the designed sequences, we measured
the mean-square distances to their respective target structures,
$\ev{\delta_0^2}$, using simulated tempering. In Tables~\ref{s_16} 
and~\ref{s_20} we give both $\ev{\delta^2_0}$ and 
$\ev{\delta^2}$ at $T=0.15$ for each of the 
sequences. From these tables a few features emerge:
\begin{itemize}
\item For target structures where the original sequence is good
(small $\ev{\delta_0^2}$), the multisequence approach either 
returns the original sequence or finds an even better sequence. 
\item For target structures where the original sequence is bad
(high $\ev{\delta_0^2}$), the multisequence approach often finds 
sequences with significantly lower $\ev{\delta_0^2}$.
\item With only one exception, structure 16-4, the results are
better or much better for multisequence design than for the energy 
minimization method. For structure 16-4, the $\ev{\delta_0^2}$ values
are relatively high for both methods, as well as for the original    
sequence.

\end{itemize}

It should be stressed that in those instances where 
the multisequence approach fails to find a good 
sequence, the original sequence is bad, too. Hence, it is likely  
that these target structures do not represent designable structures. 

While this very simple implementation of multisequence design has 
been tested with success, it should be kept in mind that there are
a number of possible improvements. As already mentioned, it would, 
for example, in off-lattice problems be more natural to maximize the fuzzy 
version of the conditional probability in Eq.~(\ref{fuzzy}), rather than 
the one referring to a single structure $r_0$ used here.

\section{Biological Implications}
\label{sec_bio}

Sequence Design, the inverse of protein folding, is of utmost 
relevance for e.g. drug design. 
The study of the statistical mechanics of protein folding is
hampered by well-known computational difficulties. In sequence
design, the major difficulty is to ensure that the designed
sequence has the target structure as its {\it global} energy
minimum. It is the ambitious goal of the multisequence design
method to achieve that, by a simultaneous search of conformation
and sequence spaces. As it stands, the method is applicable to
a fairly wide range of hydrophobic/polar models.

\section{Summary}
\label{sec_summary}

A novel MC scheme for sequence optimization in coarse-grained 
protein models has been presented 
and tested on hydrophobic/polar models.
With simultaneous moves in both 
sequence and conformation space according to a judiciously
chosen joint distribution, an efficient way of maximizing 
the corresponding conditional probabilities emerges, in which 
two different prescriptions are given for removing sequences not 
suitable for the target structures. One is
a simple energy comparison that can be applied to lattice models, 
whereas the other one is based upon the marginal distribution $P(\sigma)$
and can be applied to both lattice and off-lattice models.

The potential memory problem of keeping track of removal of most 
of the $2^N$ sequences for large $N$ is dealt with by an iterative 
method, capitalizing on the fact that the assignment of certain 
positions in the chain tend to get ``frozen'' to hydrophobic/polar 
residues. 
Furthermore, a modified algorithm was tentatively explored 
that addresses the problem of finding designable structures. This   
is highly relevant given that structures differ widely in 
designability~\cite{Li:96,Nelson:98}.

Our design method is evaluated on a number of 2D lattice ($N=16$, 18, 32 
and 50) 
and 3D off-lattice ($N=16$ and 20) structures with the following results:
\begin{itemize}
\item 
For $N=16$ and 18 lattice chains, where the results can be gauged 
against exact enumeration, the results come out extremely well both 
with respect to performance and efficiency. In this context we 
also compare with and discuss other non-exact approaches --- 
$E(r_0,\sigma)$-minimization and high-$T$ expansion. 
With respect to the former, we give, in contrast to other comparisons 
in the literature, the approach a fair chance by scanning over all 
possible net hydrophobicities. 
\item
For $N>18$ lattice chains, finding suitable design structures and 
verifying good folding properties of the designed chains is not
trivial. For $N=32$ a suitable structure was designed "by hand", 
whereas for $N=50$ a more systematic procedure was employed, where 
a variant of the multisequence approach was used to find a designable 
structure. For both $N=32$ and $N=50$ structures the results from 
the design procedure were verified to be correct.
\item 
For $N=16$ and $N=20$ off-lattice chains, a set of structures 
representing both good and bad folders were used to test the design 
method. For good folding sequences, the design procedure either 
identifies the original sequence or finds a sequence with improved 
folding properties. In the case of bad folding sequences, the design 
procedure typically finds a sequence with improved folding properties.
\end{itemize}

We also separately evaluate the efficiency of the multisequence 
approach as compared to standard MC for ordinary thermodynamic folding 
simulations. Such a test was carried out in \cite{Irback:95b}  
using carefully tuned parameters $g(\sigma)$. The results presented 
here show that it can be less expensive to fold 100--1000 chains 
simultaneously than a single one, even with a simple choice  
of $g(\sigma)$ [Eq.~(\ref{g})].

The size of the sequence optimization problem increases
rapidly with the size of the alphabet, and our approach is, as 
it stands, not practical for models with twenty amino acids. 
What might be feasible in this case is an approach along 
the lines of~\cite{Li:97}, where it was shown that an accurate 
description of the widely used Miyazawa-Jernigan $20\times20$ 
interaction matrix~\cite{Miyazawa:96} can be obtained in terms 
of its first two principal components.

\subsection*{Acknowledgement}

This work was supported by the Swedish Foundation for Strategic 
Research, the Swedish Natural Research Council and the Swedish 
Council for High Performance Computing.

\end{document}